\documentstyle[epsfig,psfig]{texas}

\begin{document}

\title{COSMIC RAYS IN CLUSTERS OF GALAXIES AND RADIO HALOS}
\author{Pasquale BLASI$^{1}$ and Sergio COLAFRANCESCO$^{2}$}
\address{(1) Department of Astronomy \& Astrophysics, The 
University of Chicago\\
5640 South Ellis Avenue, Chicago, IL 60637, USA\\
}
\address{(2) Osservatorio Astronomico di Roma\\
Via dell'Osservatorio 2, I-00040 Monteporzio-ITALY\\
{\rm Email: blasi@oddjob.uchicago.edu, cola@coma.mporzio.astro.it}}

\begin{abstract}
We calculate the fluxes of radio, hard X-rays and gamma ray emission
from clusters of galaxies, in the context of a
secondary electron model (SEM). In the SEM the radiating electrons 
are produced by the decay of charged pions in
cosmic ray (CR) interactions with the intracluster medium, while gamma 
ray emission is mainly contributed by the decay of neutral pions.
We specifically applied our calculations to the case of the Coma
cluster, and found that the combined radio and hard X-ray fluxes can 
be explained in the SEM only if very small values of the intracluster
magnetic field ($B\sim 0.1\mu G$)  are assumed, which in turn imply
a large energy density of the parent CRs. The consequent gamma-ray fluxes 
easily exceed the EGRET limit at $100$ MeV. This
conclusion can be avoided only if most of the hard X-ray emission from Coma
is not produced by Inverse Compton Scattering (ICS).

\end{abstract}

\section{Introduction}
Clusters of galaxies have recently revealed themselves as 
sites of high energy processes, resulting in a multiwavelength
emission which extends from the radio to the gamma rays and probably
beyond. In this paper we refer to the Coma cluster, because of
the wide evidences now accumulated for the presence of these 
non-thermal phenomena.

The high energy processes which produce the observable radiations
are due to the presence of a non thermal population of particles
originating most likely from the cosmic ray sources in the
cluster. The important role of cosmic ray electrons in Coma and in 
a few other
clusters of galaxies is known since a long time because of the 
diffuse radio emission which extends over typical spatial scales of order 
$\sim 1$ Mpc. This radiation can be interpreted as synchrotron 
emission of relativistic electrons in the intracluster magnetic
field. However, the combination of energy losses and diffusive
propagation of these electrons makes their motion from the sources
extremely difficult, so that it becomes a challenge to explain 
the spatial extent  of the diffuse radio emission, 
unless electrons are continuously reaccelerated in the
intracluster medium (ICM) (see \cite{feretti} for a recent
review). To solve this problem the SEM was first proposed in 
\cite{dennison,vestrand}: in this model CRs (protons) diffuse
on large scales because their energy losses are negligible and can 
produce electrons {\it in situ} as secondary products of $pp$ interactions
with the production and decay of charged pions.

The production of charged pions is always associated with the 
production of neutral pions which in turn result into 
gamma rays mainly with energies above $\sim 100$ MeV.
The flux of gamma radiation and high energy neutrinos due to the 
cosmological distribution
of clusters of galaxies was calculated in \cite{bbp} and \cite{cb}, 
where the resulting diffuse neutrino background was also evaluated.
Fluxes from single clusters were also compared with the upper limits
on the gamma ray emission from the EGRET instrument onboard the
CGRO satellite.

More recently UV \cite{lieu} and hard X-ray \cite{fusco} observations 
of the Coma cluster have led to the first detection of 
large fluxes at these wavelengths.
Thier interpretation based on ICS of relativistic electrons 
off the photons of the microwave background radiation requires an
intracluster magnetic field of $B\sim 0.1 \mu G$ \cite{fusco}.

In this paper we calculated the multifrequency spectrum of the
Coma cluster in the context of the SEM and compared our predictions 
with the results of recent observations. We find that also for 
secondary electrons, the radio and hard X-ray observations imply
an intracluster magnetic field $B\sim 0.1 \mu G$ and large
energy densities in CRs. As a consequence, the flux of
gamma rays above $100$ MeV exceeds the EGRET upper limit \cite{sreek}.
We also discuss alternative scenarios which do not imply large CR 
energy requirements.

The plan of the paper is as follows: in section 2 we describe the 
propagation of CRs in a cluster of galaxies; in section 3 we calculate
the fluxes of secondary radio, X and gamma radiation and we 
present our conclusions with application to the case of the Coma
cluster in section 4.

\section{Cosmic ray propagation in clusters of galaxies}

The propagation of CRs in clusters of galaxies was considered
in previous works \cite{bbp,volk,cb} where the effect of diffusive
confinement of CRs was investigated. We summarize these results in the following.

The CRs produced in a cluster of galaxies propagate diffusively in the
intracluster magnetic field with a diffusion time, over a spatial scale
$R$, that can be estimated as $\tau\approx R^2/4D(E)$, where $D(E)$
is the diffusion coefficient. Very little is known about the
diffusion coefficient in clusters, but as argued in \cite{bbp,cb},
for the bulk of CRs the diffusion time at large distances from the cluster center
($R\geq R_c$,
with $R_c\sim 1$ Mpc the radius of the cluster) exceeds the age of
the cluster for any reasonable choice of $D(E)$.
Assuming the following form of the diffusion coefficient,
$D(E)=D_0 E^\eta$, the maximum energy for which CRs are
confined in the cluster is
$E_c\approx (R_c^2/D_0 t_0)^{1/\eta}$,
where $t_0$ is the age of the cluster, comparable with the age
of universe. Despite of the large diffusion times in the cluster,
CR protons do not suffer appreciable energy losses in the interesting energy
range, so that the propagation can be simply
described by a transport equation with no energy loss term (see, e.g., 
\cite{blasi}).
Let us first consider the case of a single point source in the
center of the cluster, injecting CRs with a rate $Q_p(E)$.
As shown in \cite{blasi} the equilibrium number density of CRs with
energy $E$ which solves the transport equation can be written in
the form
\begin{equation}
n_p(E,r)=\frac{Q_p(E)}{D(E)}\frac{1}{2\pi^{3/2} r}
\int_{r/r_{max}(E)}^\infty dy e^{-y^2},
\label{eq:trans}
\end{equation}
where $r$ is the distance from the source and $r_{max}(E)=
\sqrt{4D(E)t_0}$. It is interesting to note that for
$r\ll r_{max}(E)$ the CR distribution tends to the well
known time independent form: $n_p(E,r)=Q_p(E)/(4\pi r D(E))$
(see \cite{bbp,cb} for details).

In the case of CRs injected homogeneously in the ICM, 
the equilibrium distribution of CRs can be written
as
\begin{equation}
n_p(E)=K\frac{\epsilon_{tot}}{V} p^{-\gamma}~,
\end{equation}
where $V$ is the volume of the cluster, $\epsilon_{tot}$ is
the total energy in CRs injected in the cluster and we
assumed that the injection spectrum is a power law in the
CR momentum $p$. The constant $K$ is a normalization constant
determined by energy conservation. Clearly this solution
breaks down close to the boundary of the cluster and at high
energies, where CRs are no longer confined in the cluster.

\section{Cosmic ray interactions and secondary electron emission}

The main interaction channel of CR protons in clusters of
galaxies is represented by $pp$ collisions with pion production.
The decay of neutral pions produces gamma rays with
energy above $\sim 100$ MeV, while the decay of charged pions
results in electrons and neutrinos. The production of gamma rays and
neutrinos from clusters of galaxies was recently investigated in
\cite{bbp,cb}. The secondary electrons produced by charged pions
can play a fundamental role in the explanation of not thermal
emission in clusters of galaxies, and here we describe
this point in a greater detail.

The `primary electrons' models proposed as an 
explanation of radio halos in Coma-like clusters 
have serious problems due to the severe energy losses that make
difficult the propagation of the relativistic electrons out to Mpc
scales, where the diffuse radio halo emission is observed.
This problem would be solved if electrons were produced
or accelerated {\it in situ}. As initially proposed in
\cite{dennison,vestrand} this is the case for
secondary electrons, generated in
CR interactions with the thermal gas in the ICM.
The same electrons would also produce X-rays by inverse compton
scattering (ICS) off the photons of the microwave background.

The calculation of the production spectrum of secondary electrons
is explained in detail in \cite{blasi}, and will be summarized
here. For both radio and X-ray emission, the relevant electrons
have energies above $\sim 1$ GeV, and it is important to
have a good description of the $pp$ interaction in a wide
range of energies. For $pp$ collisions at laboratory energy less
than $\sim 3$ GeV, the pion production is well described by a
isobar model, in which the pions are the result of the decay of
the $\Delta$ resonance \cite{dermer,strong}. For energies larger
than $\sim 7-10$ GeV we use a scaling approach. In the latter
the cross section for $pp$ collisions depends only on the ratio
of the pion energy $E_\pi$ and the incident proton energy $E_p$
($x=E_\pi/E_p$) and can be written in the following form
\begin{equation}
\frac{d\sigma(E_p,E_\pi)}{dE_\pi} = \frac{\sigma_0}{E_\pi}
f_\pi(x) ~,
\end{equation}
where $f_\pi(x)$ is the scaling function given in \cite{blasi} for
the case of charged and neutral pions. We refer to \cite{strong}
for a detailed expression of the cross section in the low energy
case.

The production electron spectrum at distance $r$ from the
cluster center, assumed to be spherically symmetric,
can be easily calculated according to the expression
\begin{eqnarray}
q_e(E_e,r) &=& \frac{m_{\pi}^2}{m_{\pi}^2-m_{\mu}^2} n_H(r) c \cdot
\int_{E_e}^{E_p^{max}} dE_\mu \int_{{E_\pi}^{min}}^{{E_\pi}^{max}}
dE_\pi 
\nonumber \\
& &
\int_{E_{th}(E_\pi)}^{E_p^{max}} dE_p~F_\pi(E_\pi,E_p)
F_e(E_e,E_\mu,E_\pi) n_p(E_p,r) ~,
\label{eq:source}
\end{eqnarray}
where $F_\pi$ is the differential cross section for the production
of a pion with energy $E_\pi$ in a $pp$ collision at energy $E_p$
(see \cite{blasi} for details),
$F_e$ is the spectrum of electrons generated by the decay of a
single muon with energy $E_\mu$ and $n_p$ is the spectrum of CRs.
The function $F_e$ depends also on the pion energy because the
muons produced in the pion decay are fully polarized, and this
effect is taken into account here.
The gas density at distance $r$ is assumed to follow a King profile
\begin{equation}
n_H(r)=n_0\left[ 1+\left(\frac{r}{r_0}\right)^2\right]^{-3\beta/2},
\end{equation}
where, in the case of the Coma cluster we use $n_0=3\times 10^{-3}
cm^{-3}$, $r_0=400$ kpc and $\beta=0.75$ (we use here $H_0=60 ~km ~ s^{-1} ~
Mpc^{-1}$).

The equilibrium electron distribution, $n_e(E_e,r)$,
is achieved mainly due to
energy losses, dominated at high energy by ICS and synchrotron
emission and at low energy by Coulomb scattering.
The effect of losses is to produce a steepening of the electron
spectrum by one power in energy, at high energy, and a flattening
by one power of energy, at low energy.
In the next subsections we outline the calculations of non thermal
radio, X and gamma ray emission from a cluster.

\subsection{The radio halo emission}

Electrons with energy $E_e$ in a magnetic field $B$ radiate
by synchrotron emission photons with typical frequency
\begin{equation}
\nu=3.7\times 10^6 B_\mu E_e^2 Hz,
\label{eq:nu}
\end{equation}
where $B_\mu$ is the value of the magnetic field in $\mu G$.
The emissivity at frequency $\nu$ and at distance $r$ from the cluster
center can be easily estimated as
\begin{equation}
j(\nu,r)=n_e(E_e,r)\left( \frac{dE_e}{dt}\right)_{syn}
\frac{dE_e}{d\nu},
\label{eq:jnu}
\end{equation}
where $(dE_e/dt)_{syn}$ denotes the rate of energy losses due to
synchrotron emission and $dE_e/d\nu$ is obtained from
eq. (\ref{eq:nu}).
The total fluence from the cluster is obtained by volume integration.

Some simple comments can help understanding general features of
the radio halo spectrum: for this purpose let us assume that CR are
confined in the cluster and that the density of intracluster
gas is spatially constant.
If the injected CR spectrum is
$\propto E_p^{-\gamma}$ then the equilibrium CR spectrum is,
within the distance $r_{max}(E)$ defined above, a power law
$\propto E_p^{-(\gamma+\eta)}$.
Provided the electron energy is $E_e\geq 1$ GeV, the production
electron spectrum reproduces the parent CR spectrum, so that
the equilibrium electron spectrum in the same energy region is
$\propto E_e^{-(\gamma+\eta+1)}$. From eq. (\ref{eq:jnu}) it
is easy to show that
$j_\nu(\nu,r)\propto \nu^{-(\gamma+\eta)/2}$. The volume integration
gives the observed spectrum: in the simple assumptions used here
(complete confinement)
the integration over the distance $r$ at each frequency $\nu$
must be limited by a maximum value $r_{max}(\nu)\propto \nu^{\eta/4}$,
so that the resulting radio halo spectrum is $\propto \nu^{-\gamma/2}$,
independent of the diffusion details.
Repeating the same discussion, it is easy to show that
the synchrotron radiation produced by CRs non confined in the
cluster volume has a spectrum as steep as $\nu^{-(\gamma+\eta)/2}$. Therefore
there is, in principle, a break frequency where the spectrum steepens
from a power index $\gamma/2$ to a power index $(\gamma+\eta)/2$.
In any realistic scenario, there is a smooth steepening
of the spectrum. The presence of a density profile in the
intracluster gas also contributes an additional small steepening of the spectrum.
All these propagation effects are self-consistently considered 
in eq. (\ref{eq:trans}).

\subsection{Non thermal X-rays}

The peak energy where most of the photons are produced by ICS
of electrons with energy $E_e$ is
\begin{equation}
E_X=2.7 E_e^2(GeV) keV~.
\label{eq:Ex}
\end{equation}
So, that the emissivity in the form of X-rays with energy $E_X$
at distance $r$ from the cluster center is:
\begin{equation}
\phi_X(E_X,r) = n_e(E_e,r) \left( \frac{dE_e}{dt}\right)_{ICS}
\frac{dE_e}{dE_X}.
\end{equation}
Here $(dE_e/dt)_{ICS}$ is the rate of energy losses due to ICS
and $dE_e/dE_X$ is calculated from eq. (\ref{eq:Ex}).
As for the radio halo emission, the observed fluence is determined by
volume integration.

\subsection{The gamma ray emission}

As pointed out above, the production of neutral pions in
$pp$ collisions results in the generation of gamma rays
with typical energy $E_\gamma\geq 100$ MeV.
The emissivity of gamma rays with energy $E_\gamma$ at
distance $r$ from the center of the cluster is given by
\begin{equation}
j_\gamma(E_\gamma,r)=2 n_H(r) c
\int_{E_\pi^{min}(E_\gamma)}^{E_p^{max}}
\int_{E_{th}(E_\pi)}^{E_p^{max}}
dE_p F_{\pi^0}(E_\pi,E_p) \frac{n_p(E_p,r)}
{(E_\pi^2+m_\pi^2)^{1/2}},
\end{equation}
where $E_\pi^{min}(E_\gamma)=E_\gamma+m_{\pi^0}^2/(4E_\gamma)$.
The function $F_{\pi^0}$ is again calculated using the isobar model 
for proton energy less than $3$ GeV and with the scaling model for energies 
larger than $7$ GeV.
The observed gamma ray flux is obtained by integration
over the cluster volume.

In \cite{blasi} the contribution of secondary electrons to the
gamma ray flux through bremsstrahlung was also calculated. Since
its contribution is small if compared with the contribution from
pion decay, we neglect here the bremsstrahlung gamma ray emission.

\section{Application to the Coma cluster}

In this section we apply our predictions in the SEM framework 
to the case of the Coma cluster, for which
multiwavelength observations are available.

Two extreme scenarios of CR injection are considered here: a point-like CR source
in the cluster center and a uniform CR injection distributed over the
cluster volume. In both cases we assume that the CR injection spectrum
is a power law in momentum with power index $2.1\leq \gamma\leq 2.4$,
covering the range of values expected for first order Fermi
acceleration at shocks as well as for other CR acceleration mechanisms.

The diffusion of CRs in the cluster is described by a diffusion
coefficient derived from a Kolmogorov spectrum of magnetic
fluctuations in the cluster, according with the procedure described
in \cite{cb}. 
In this case the diffusion coefficient can be written as
\begin{equation}
D(E)=2.3\times 10^{29} E(GeV)^{1/3} B_\mu^{-1/3}
\left(\frac{l_c}{20 kpc}\right)^{2/3}~cm^2/s
\end{equation}
where $l_c$ is the size of the largest eddy in the Kolmogorov
spectrum.

The procedure used to evaluate the expected fluxes is
the following: we first fit the spectrum of the radio halo 
as given in \cite{radio} for $\gamma=2.1$ and $\gamma=2.4$.
This allows to find the value of the absolute normalization of the
injection CR spectrum in terms of an injection luminosity $L_p$
as a function of the average intracluster magnetic field $B_\mu$.
We carried out this calculation for $B_\mu=0.1,1$ and $2$. After calculating
$L_p$ in this way, we determine the hard X-ray flux and the gamma ray
flux above $100$ MeV according to the expressions
given in the previous section. Our results for the radio
and hard X-ray emission for the case of a single source are shown in
Figs. 1 and 2 respectively. 
The three panels 
refer to values $B_\mu=0.1,1,2$, as indicated. The normalization
constants and the integral fluxes of gamma rays above $100$ MeV
in the same cases are reported in Table 1, where the
gamma ray flux is compared with the upper limit obtained by the
EGRET experiment \cite{sreek}. The last column in Table 1 contains the
ratio of the flux of gamma rays from electron bremsstrahlung compared 
with the flux of gamma rays from pion decay, as calculated in \cite{blasi}.

\begin{figure}
\begin{center}
\psfig{file=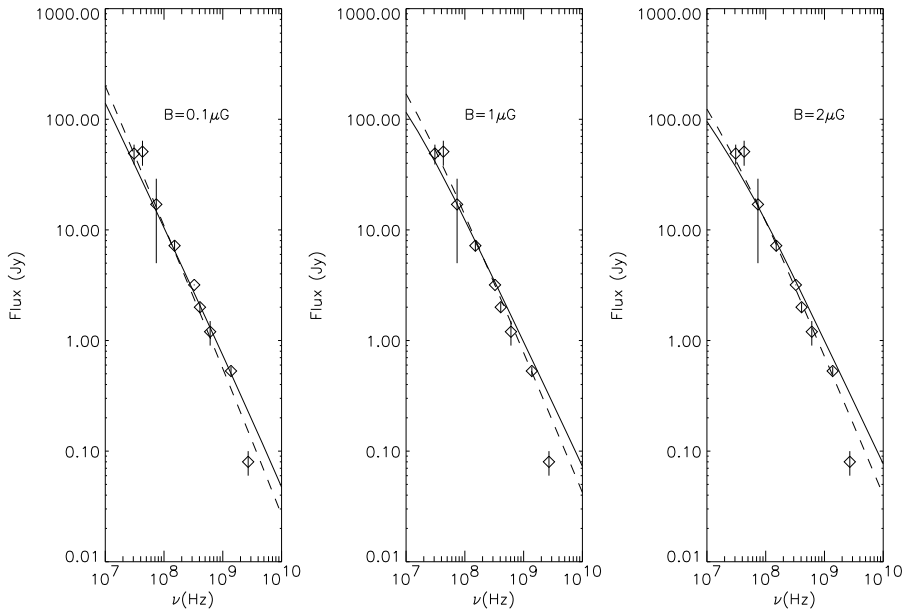,width=12cm}
\caption{Fluxes of radio radiation from the Coma cluster 
calculated in the SEM for
$\gamma=2.1$ (solid lines) and $\gamma=2.4$ (dashed lines). The three
panels are, from left to right, for $B_\mu=0.1,~1,~2$.}\label{fig1}
\end{center}
\end{figure}

Fig. 2 shows that a joint fit for the radio and hard X-ray fluxes
is possible only for low values of the magnetic field ($B_\mu\sim 0.1$),
which in turn imply large values of $L_p$ (see Table 1). As a
consequence, the gamma ray fluxes easily exceed the EGRET upper
limit from Coma. These results indicate that the SEM cannot explain
the multiwavelength observations of the Coma cluster without
violating the EGRET limit. This conclusion is not appreciably
changed in the case of homogeneous injection of CRs. In this case,
the value of $\gamma$ is fixed by radio observations 
and is
$\gamma=2.32$. In order to fit the hard X-ray data at the same time,
an ICM magnetic field $B_\mu\simeq 0.1$ and a total energy in CRs 
$\epsilon_{tot}\approx 8\times 10^{63}$ erg are required. 
This value, averaged
over the age of the cluster, corresponds to a typical CR luminosity
of $L_p\approx 2\times 10^{46}$ erg/s, which implies a gamma ray 
flux above $100$ MeV in excess of the EGRET upper limit by a factor 
$\sim 3$.
Therefore, also for a  homogeneous CR injection,  the SEM fails
in explaining the radio and hard X-ray observations at the same time,
without exceeding the EGRET limit.

\begin{figure}
\begin{center}
\psfig{file=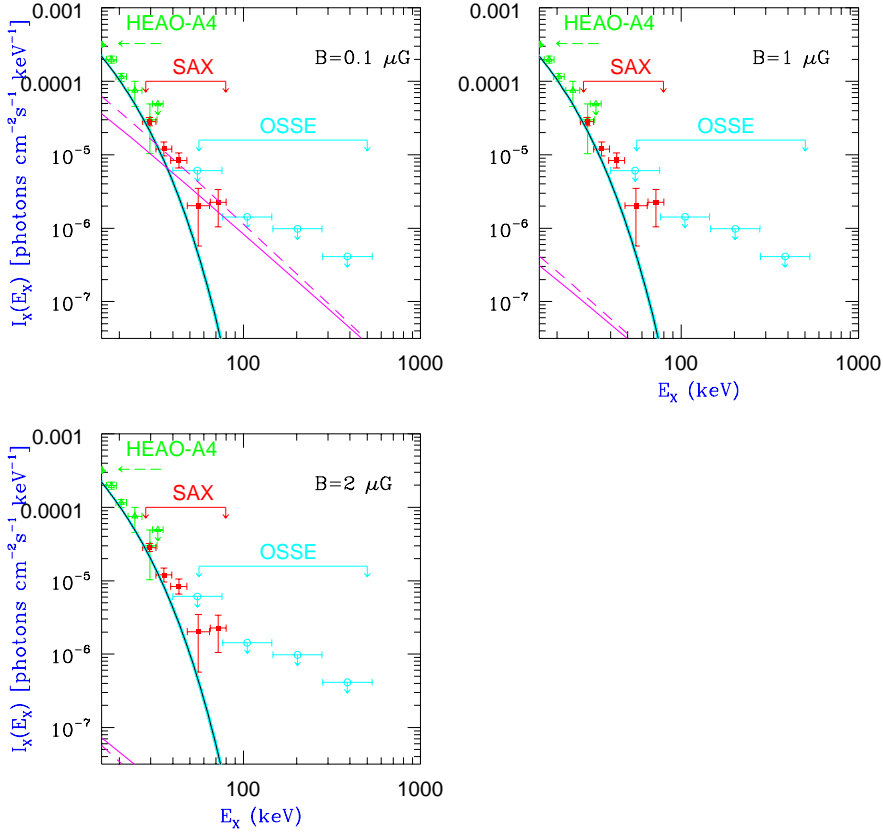,width=12cm}
\caption{Spectrum  of the diffuse X-ray emission from Coma.
The three panels refer to the same values of the IC magnetic field as
in Fig. 1.
The shaded area shows the best fit to the HEAO1-A4 and GINGA
thermal emission data (open triangles) at $T=8.21 \pm 0.20$ keV [18].
The OSSE upper limits [17]
are indicated by the open circles.
The SAX data [7]
are indicated by filled squares.
For this data set, errors are shown at $90 \%$ confidence level.
Arrows and labels show, for each panel, the energy ranges in which the
three different data sets are located.
Predictions of the SEM for $\gamma=2.1$ (solid lines) and
$\gamma=2.4$ (dashed lines) are shown in each panel.
}
\label{fig2}
\end{center}
\end{figure}

This conclusion has, however, further important consequences for 
other models too. In fact, we checked that the 
CR energy densities obtained in the present paper are comparable
with those obtained assuming equipartition, as done in recent papers
(e.g. \cite{Lieuetal})
to explain the radio, hard X-ray and UV observations. 
Therefore, the gamma ray limit applies to other models as well
and forces the CR energy density in clusters to be some fraction
of the equipartition value. Actually, as shown in \cite{apjlett},
the present gamma ray observations put weak constraints on this 
fraction, but future gamma ray observations will definitely do better.
We can envision at least two other arguments against 
equipartition of CRs and the  ICM in clusters of galaxies: first of all,
the most powerful CR sources typically present in clusters of
galaxies (see \cite{bbp,cb}) allow to achieve a CR energy density equal 
to a small fraction (typically $\sim 5\%$) of the equipartition value.
Moreover, the magnetic field derived in \cite{fusco} ($B\sim 0.1 \mu G$),
which is the main reason to call for equipartition, is quite smaller
than the equipartition magnetic field in the cluster, and it seems
difficult to envision a scenario where CRs are in equipartition but
not magnetic fields, in particular if the origin of CRs and magnetic
fields in clusters are related each other.

{\Large
\begin{center}
\begin{table}
\caption{Summary of the fitting parameter values
\label{table1}}
\vskip 0.3truecm
\begin{tabular}{| c | c | c | c | c |}
\hline
\hline
$B_\mu$ & $\gamma$ & $\frac{L_p}{10^{44}erg s^{-1}}$ &
$\frac{F_\gamma(E_\gamma\geq 100MeV)}{F_\gamma^{EGRET}
(E_\gamma\geq 100MeV)}$ & $F_{\gamma}^{brem}/F_{\gamma}^{\pi^0}$
\\ \hline

$0.1$ & $2.1$ & $50$ & $1.93$ & 0.13\\
$0.1$ & $2.4$ & $180$ & $7.15$ & 0.10 \\
$1$ & $2.1$ & $0.35$ & $1.8\cdot 10^{-2}$ & 0.14 \\
$1$ & $2.4$ & $1$ & $4.5\cdot 10^{-2}$ & 0.11\\
$2$ & $2.1$ & $0.1$ & $5.3\cdot 10^{-3}$  & 0.12\\
$2$ & $2.4$ & $0.23$ & $1.1\cdot 10^{-2}$ & 0.095\\

\hline
\hline
\end{tabular}
\end{table}
\end{center}
}

So, at the present status of the debate, it is necessary to look for 
possible 
alternative explanations of the hard X-ray excess observed in Coma with 
the SAX satellite, since there is no stringent
argument in favour of the ICS origin of such hard X-ray tail.
A possible alternative was proposed 
in \cite{ensslin} where it was speculated that the presence 
of a non thermal tail in the electron distribution could account
for the X-ray excess through bremsstrahlung emission. 
If this turns out to be the explanation of the hard X-ray
excess, then the radio and hard X-ray fluxes become unrelated and 
magnetic fields of order $\geq 1 \mu G$,
as the ones suggested by Faraday rotation measurements,
would still be allowed. As a consequence,
the CR energy density required to fit the radio and hard X-ray data
would be of the same order of that predicted
in \cite{bbp,cb}. In this case, the 
SEM still remains a viable option for the origin of the Coma radio 
halo emission.

\end{document}